\documentclass[preprints,article,accept,moreauthors,pdftex]{Definitions/mdpi} 
\firstpage{1} 
\pubvolume{1}
\issuenum{1}
\articlenumber{0}
\pubyear{2021}
\copyrightyear{2020}
\datereceived{} 
\dateaccepted{} 
\datepublished{} 
\hreflink{https://doi.org/} 

\newcommand{\tg}{\#{\em tag~ }}
\newcommand{\tgs}{\#{\em tags~ }}
\newcommand{\Tgs}{\#{\em Tags~ }}
\Title{Radio Galaxy Classification:  \#Tags, not Boxes}

\TitleCitation{\#Tags, not Boxes}

\Author{Lawrence Rudnick $^{1}$}

\AuthorNames{Lawrence Rudnick}

\AuthorCitation{Rudnick, L.}

\address{%
$^{1}$ \quad University of Minnesota; larry@umn.edu\\
}

\corres{Correspondence: larry@umn.edu; Tel: 1-612-624-3396}

\firstnote{Current address: Minnesota Institute for Astrophysics, University of Minnesota, 116 Church St., SE, Minneapolis, MN 55455, USA} 

\abstract{After six decades of studying radio galaxies, we are now being delightfully overwhelmed by their exponentially expanding numbers, and the complexity of their structures. Similarly, the ways we classify radio galaxies have exploded, often leading to conflicting terminology, ambiguous classifications, and historical schemes that may or may not match with our current physical understanding.  After discussions with more than 100 radio astronomers over the last several years, listening to their ideas and aspirations, I propose that we reconceptualize the classification of radio galaxies. Instead of trying to put them into "boxes",  we should assign them  \tgs, a system that is easy to understand and apply, flexible and evolving, and can accommodate conflicting ideas about what is relevant and important.  Here, I outline the basis of such a \tg system; the rest is up to the community. }


\begin{document}
\section{Introduction}
Why do we classify radio galaxies, or any objects for that matter?  I see three main benefits to classification, in general:
\begin{itemize}
    \item {Communication -- so we have language to describe what objects we are interested in. {\em "I'm investigating one-sided jets."}}
\item{Investigation -- so we can define samples for study, e.g. to look for correlated variables. {\em "I'm studying whether one-sided jets appear more frequently in variable core sources."}}
\item{Interpretation -- so we can develop physical models to explain observed phenomena. {\em "I'm working on how relativistic beaming can determine whether jets would appear one- or two-sided."}} \end{itemize}

The challenges inherent in our classification system are well-known, but difficult to avoid.  Different people often mean different things but use the same term  - could we get a group of astronomers to decide on what a {\em dying radio galaxy} is?   Often, objects don't fit nicely into the boxes we've defined -- ADS lists 230 articles with "hybrid", "radio" and "galaxy" in their abstracts. And in a remote talk for the 2019 3Csky conference in Turin, Bernie Fanaroff (of FR reknown) lamented "{\em [recently] I worked on a sample of sources.  I couldn't classify them. They were too ambiguous and irregular.}"   In other cases, we group together objects with very different physical origins, (e.g., the broad variety of "X-shaped" radio galaxies) confounding studies that look for correlations, develop physical models, etc.  

So where are we today?  Table \ref{tab1} summarizes many of the classes of sources in the current literature.  Some of these classes are exclusive, i.e., a source can be one or the other, e.g., either ``flat" or ``steep" spectrum, and either ``S-shaped" or ``C-shaped", but not both.  Other classes can be combined, e.g., a source can be in both the FRI and narrow-angle tail classes. These multiple uses,  and the lack of a clean definition for many of the classes, can lead to confusion, and worse, can lead to scientifically inconsistent results from different studies.   Other measured quantities, such as polarization information as a function of wavelength, or X-ray counterparts, add additional layers of complexity.
\begin{specialtable}[H] 
\small
\caption{Examples of radio galaxy classifications in current use.\label{tab1}}
\begin{tabular}{ccc}
\toprule
\textbf{Category}	& \textbf{Classes}	\\
\midrule
Morphology	& Double; Classical double; Triple; Narrow-angle tail; Wide-angle tail; \\ 
 & Bent-tail; FRI, FRII, FR0*; Hybrid; X-shaped;\\
&  S-shaped; C-shaped; Relaxed; Dying RG;  Double-double; \\
& Core-dominant; Core-halo; Core-jet; CSO; 1-sided  \\
 Size & Compact (pc); Galactic (<10 kpc); extended RG (10-1000 kpc);\\
   & Giant RG (>1 Mpc)\\
Host 	& Radio Galaxy; SFG: Spiral; Seyfert I,II;  QSO; Blazar  \\
  & BLLac; BLRG; NLRG; ULIRG; LERG; \\
  & HERG; LINER; BCG	\\
  Spectra		& Flat; Steep; Ultra-steep; GigaHz Peaked; \\
  & Inverted; Convex; Concave; Complex			\\
\bottomrule
\end{tabular}
\\ *This unfortunate nomenclature is now being more widely adopted and should by dropped;  see the critique in \citet{hard20}.
\end{specialtable}
Over the past several years, I have led discussions and surveys on these issues with over 100 members of the radio astronomy community, at meetings and workshops; I acknowledge without mentioning names the many people who contributed fertile ideas.  Several lessons became clear:
\begin{itemize}
    \item {The current classification schemes are confusing and have started to break down.}
    \item{As a community, we want to document a very wide variety of characteristics for large samples of radio sources. Some of the more commonly mentioned (at least today) are listed in Table \ref{priorities}.}
    \item{Multi-wavelength information should be included where available.}
    \item{Confidence levels should be provided.}
    \item{Classifications should evolve as more information becomes available.}
    \item{Criteria for classification should be completely transparent.}
\end{itemize}

\begin{specialtable}[H] 
\small
\caption{Commonly mentioned priorities for catalog {\em source} descriptions. These assume that {\em components} have been assembled into {\em sources}. Some require information external from the survey.} \label{priorities}
\begin{tabular}{ccc}
\toprule
\textbf{Category}	& \textbf{Measurements/Descriptors}	\\
\midrule
Direct & Peak and total flux; Brightness temperature;\\
 & Angular size and area; Linear Size and area;\\
  & Redshift; Spectral index; Fractional polarization; RM;\\
  & Number of components/peaks in source\\
 Structural & Core/total flux;  Number of jets (0,1,2); Jet-flux/total-flux;\\
  & Peak-separation/total-extent (FRI,II); Shape (e.g., linear, bent);\\
  & Symmetry (S-, C-, X-)\\
  Supplemental & Host properties (include SFR); X-ray properties;\\
   & pc-scale structure; group or cluster environment\\
 
 \bottomrule
\end{tabular}
\end{specialtable}

In the face of these challenges and aspirations,  we are fortunately at a time of great opportunity, since new catalogs with millions of radio components are being or will be produced by LOFAR,  EMU and POSSUM at ASKAP, MeerKAT, VLASS at the VLA,  Apertif at Westerbork, $\mu$GMRT, GLEAM at the MWA, etc..  If new classification schemes can be incorporated into those catalogs, then their scientific usefulness will blossom.

\section{\#Tags}

To take advantage of these opportunities, I propose that instead of putting a source into a {\em box}  so that it belongs in that box and not in other boxes, we use a series of criteria-based \tgs.  Each source can have any number of \tgs, as long as the source passes the criteria for each \tg.  

This system forces a subtle, but important distinction in how we think about these classes  --  instead of deciding  that  "Source A {\em IS} a member of Class X,"  we say, more narrowly,  "In Survey G, Source A passes the test to be assigned   \#{\em X}."   Although this may seem trivial, it provides enormous advantages.  For example, no longer will we need to argue about whether a source detected in a high-resolution survey is {\em really}  a small one-sided jet; if it passes the criteria, it will be assigned \#{\em one-sided-jet}.  Then,  if a new low frequency survey shows that it is part of a much larger structure, with two jets and extended lobes visible, then the new survey will assign it a different \tg, particular to that survey.  \Tgs are thus explicitly dependent on the properties of the observations/survey that were used to make the assignment.

\begin{figure}[H]
\begin{center}
\includegraphics[width=10.5 cm]{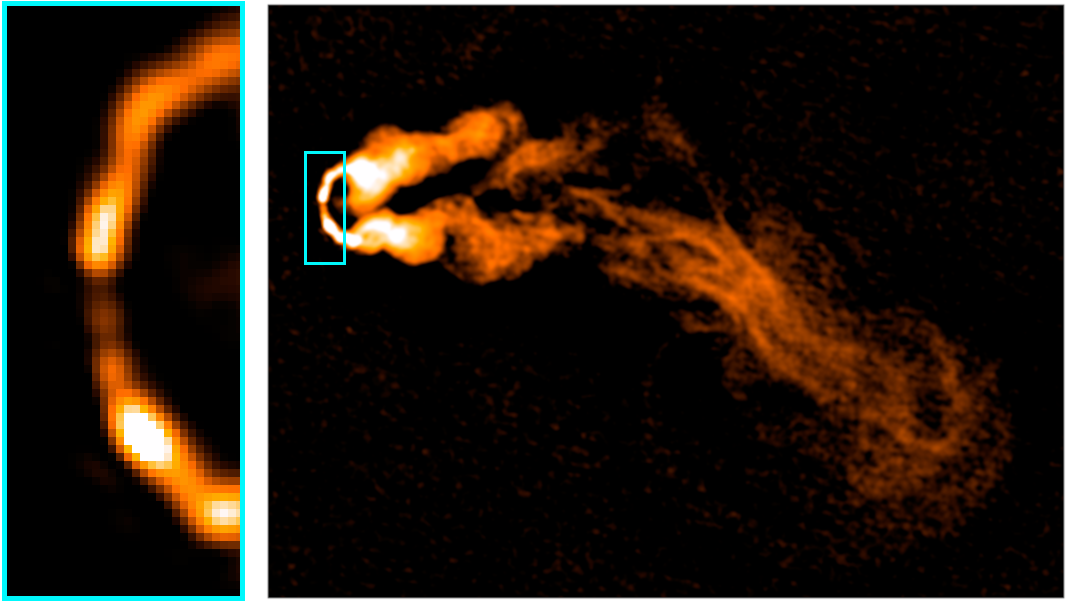}
\caption{ \textbf{Is this not a WAT?}. Using the VLA data presented in \citet{MLpers}: \textbf{(Left)} Close-up view  of the head of NGC~1265;   \textbf{(Right)} Larger scale view of the same source.}
\label{1265}
\end{center}
\end{figure}

 Figure \ref{1265} gives a further example of this re-conceptualization. It shows two views of the prototypical narrow-angle-tail source NGC~1265.   Such tails, when seen near the AGN at high resolution, will always look like wide-angle-tails (WATs).  If we follow the criteria in the defining paper for WATs (\citet{WAT}), the ``source" on the left would be classified a \#{\em WAT}.  This is not a mistake --  if we follow the \tg schema; it simply means that it passed the relevant test.  A theorist, e.g., may well want  to model the properties of this source as part of studying the gently bent portion of jets, and select it as part of a sample of \#{\em WAT}s from a catalog.   On the other hand, an observer studying ICM interactions might want to exlcude this \#{\em WAT} based on other catalog entries, e.g., because its small linear size shows that it is still being influenced by its host's ISM.

\subsection{\#{\em Tag} Principles}

The following summarizes the principles, characteristics of \tg systems that are required to meet the scientific needs expressed in our discussions.  These are likely to need refinement once such systems go into use.

\begin{itemize}
\item{\Tgs {as proposed here} apply to {\em sources}, where all the emission is believed to originate in a single host, whether visible or not. \Tgs {currently} do not apply to the source's constituent components, as identified by ``source" finders, {although such schemes could also be developed}.}
    \item {Each source can have multiple \tgs. }
    \item {A source can pass or fail to meet the criteria of each \tg, or not have sufficient information to be tested. For example, a barely resolved double source cannot be tested for the number of jets present.}
    \item {There can be multiple versions of \tgs for the same purpose, with different underlying criteria. For example, \#{\em Giant$_A$}  might include bent sources, where the sum of the length of the two lobes was $>$700~kpc, while \#{\em Giant$_B$} might only include sources that cannot fit in a 700~kpc box.}
    \item{\Tgs must be based on well-defined criteria, quantitative wherever possible, with the definitions or algorithms made available.}
    \item{\Tgs can change with time as more information or better algorithms become available.  (Keeping track of versions will become important.)}
    \item{\Tgs should have corresponding confidence values, if possible.}
    \item{\Tgs will be specific to a given survey, since they will depend on resolution, sensitivity, dynamic range, availability of auxiliary information, etc.}
    \item{\Tgs can be valuable even when they appear trivial.  For example, although sizes and errors may be included in a catalog, the distinction between extended and compact sources may need expert judgement, and be embodied in the \tg.  The experts would consider the dependencies on signal:noise, dynamic range, the presence of artifacts, etc. in setting up the \tg criteria.}
\end{itemize}

\subsection{Tag Example}
The following illustrates the application of \tgs  to an anonymous source from the MeerKAT Galaxy Cluster Legacy Survey (\citet{knowles21}).  Only a handful of \tgs are shown;  each catalog may use as many \tgs as are appropriate or practical.  Note that there will be many other catalog entries for this source.  These would include positions, fluxes, etc as listed in Table \ref{tab1}, and would be used as criteria for each \tg.   The assumption, for this illustration, is that in-band spectral indices are available,  as well as the optical host, but not a redshift.
\begin{figure}[H]
\begin{center}
\includegraphics[width=5 cm]{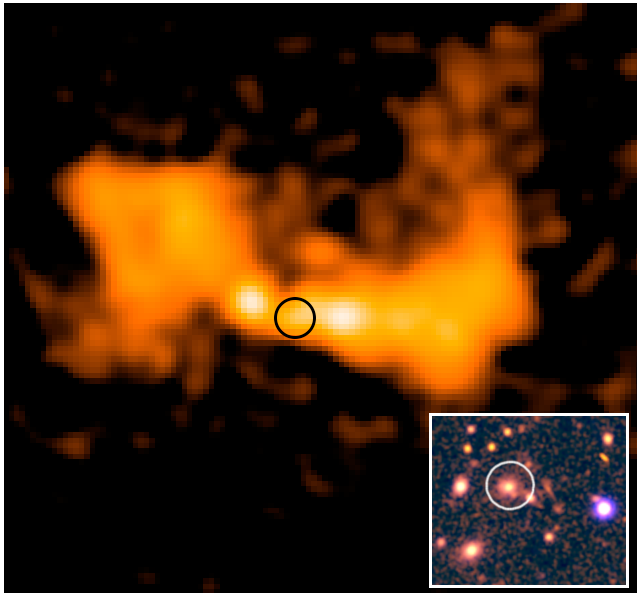}
\caption{ Anonymous example source from the (fictitious) EMK Survey, for  the application of \tgs. Radio map is actually from \citet{knowles21}.  Optical inset is from the Dark Energy Survey.}\label{examp}
\end{center}
\end{figure} 

\begin{specialtable}[H] 
\small
\caption{\#{\em Tag} example for source in Figure \ref{examp}. +1 = satisfied \tg criteria.   -1 = failed \tg criteria.  0~=~insufficient information to test \tg criteria. Each \tg is prepended by EMK, to identify it with the (fictitious) EMK survey.}
\label{tab2}
\begin{tabular}{cc|cc}
\toprule
\textbf{{\em \#Tag}}	& \textbf{Value} &\textbf{{\em \#Tag}}	& \textbf{Value}	\\
\midrule
\ul {Structural} &  & \ul {Spectral+}& 		\\
\#{\em EMK:S-symmetry}          & +1     & \#{\em EMK:SteepSpec}     & +1\\
\#{\em EMK:C-symmetry}          & -1    & \#{\em EMK::ConvexSpec}   & 0 \\
\#{\em EMK:FRI} (peaksep/tot$<$0.5)     & +1    &  \#{\em EMK:Backflow} & -1 \\
\#{\em EMK:FRII} (peaksep/tot$>$0.5)    & -1    & \#{\em EMK:Outflow}   &  +1 \\
\#{\em EMK:Coredom}             & -1      & \#{\em EMK:Polarized} & 0\\
\#{\em EMK:Giant}               & 0     & \#{\em EMK:BCG}       & 0\\

\bottomrule
\end{tabular}
\end{specialtable}
Careful readers will note two unfamiliar \tgs, \#{\em EMK:Backflow} and \#{\em EMK:Outflow}.  I introduce them here as an example of how individuals can add \tgs that they think are important, without affecting earlier alternative \tgs.  In this case, the new \tgs are alternatives to the FRI,II scheme, and are likely to be much more scientifically informative. \#{\em EMK:Backflow} denotes sources with spectral indices that steepen from their leading edge back to the nucleus;  \#{\em EMK:Outflow} sources have spectral indices that steepen with increasing distance from the nucleus.  These represent two physically different jet behaviors, and the FRI,II scheme can be understood as a less precise proxy for these behaviors. The disadvantage of these new \tgs is that they can be applied only to sources that are sufficiently large and bright; but since they only add information, nothing is lost.  Another technical detail is that spectral changes due to magnetic field variations on a curved electron spectrum would have to be considered in applying this criterion, to isolate the effects of radiative ageing.

\subsection{ Implementation of \Tgs}
Although individual investigators could and should develop their own \tgs to address questions of scientific interest, the most useful use of \tgs will come from observatory or survey teams that incorporate them into catalogs. The EMU Survey has begun such a process in the design of {\em EMUcat}, with many of the details still being worked out. Following  are some notes and guidelines that may be useful in implementing such a scheme.  As before, these will certainly need to be revisited as the community gets experience with this new classification tool. 
\begin{itemize}
    \item {{For a source-based \tg scheme,} components from source-finders will first have to be assembled into sources, with or without a host identification.}
    \item {Each \tg will require measurements or other data about the source, that are already included in the catalog.  Early identification of \tgs to be used will ensure that the required information is available.}
    \item {Some \tgs will be based on catalog data; other \tgs will require algorithms to be run on images (e.g., symmetries, jet presence or dominance)}
    \item{Some \tgs will require information from auxiliary databases not included with the catalog; links or other references to that information are needed.}
    \item{All criteria and algorithms should be made publicly available, either as metadata, or in software repositories, etc.}
    \item{Different surveys will require different schemes;  however, wherever common definitions can be used, those will be highly desirable.}
    \item{It is desirable to allow for  new \tgs developed by individuals and teams to be added to the catalog, after appropriate vetting by the survey team.}
    \item{Versioning of \tgs will have to be built into the catalog design.}
\end{itemize}

{One quite promising opportunity and challenge to a strict algorithmic approach is to} incorporate classifications that are done by visual inspection, such as Radio Galaxy Zoo (\citet{rgz}) or LOFAR Galazy Zoo (\citet{lzoo}) and their planned successors.  The use of MaNGA morphologies from Galaxy Zoo  for SDSS entries (\citet{manga}), where available, may provide guidance on how to include non-quantitative \tgs.  Similarly, classifications from supervised and unsupervised machine learning algorithms will become available over the next several years, and the community will have to determine how to best utilize those. One innovative effort using machine learning to build on a \tg-like schema has now been used in the classification of supernova spectra (\citet{sn21}).

\section{Conclusions:}
The mammoth new catalogs and increasingly sophisticated scientific questions being asked about radio galaxies demand a new approach to radio galaxy classification.  \Tgs provide a promising alternative.\\

\acknowledgments{Although I've been interested in this issue for many years, I credit Ray Norris for ``recruiting" me to take on this task because of the demands imposed by the EMU Survey. The advice of many colleagues over the last several years has been of immeasurable value.  The questions surrounding implementation have benefited enormously by the work of Josh Marvil and his EMUcat team.}

\funding{This work was supported, in part, by U.S. National Science Foundation grant AST17-14205 to the University of Minnesota.}

\conflictsofinterest{The author declares no conflict of interest.}
\reftitle{References}



\begin{thebibliography}{999}
\bibitem[Banfield et al. (2015)]{rgz}
Banfield, J. K., Wong, O. I., Willett, K. W. et al. Radio Galaxy Zoo: host galaxies and radio morphologies derived from visual inspection, {\em MNRAS}, {\bf 2015}, {\em 453}. 2326.
\bibitem[Davison, Parkison, Tucker (2021)]{sn21} 
STag: Supernova Tagging and Classification . Davison, William,  Parkinson, David,  Tucker, Brad E. {\em arXiv2 108:10497} {\bf 2021}
\bibitem[Gendron-Marsolais et al. (2020)]{MLpers}
Gendron-Marsolais, M.,  Hlavacek-Larrondo, J., van Weeren, R. J., Rudnick, L. et al., {\em MNRAS}, {\textbf 2020}, {\em 499}, 5971.
\bibitem[Hardcastle and Croston (2020)]{hard20}
Hardcastle, M. and Croston, J., Radio galaxies and feedback from AGN jets,  {\em NewAR}, {\textbf 2020}, {\em 88}, 101539.
\bibitem[Knowles et al. (2021)]{knowles21}
Knowles, K., Cotton, W., Rudnick, L. et al.  The MeerKAT Galaxy Cluster Legacy Survey, submitted to {\em A\&A}, {\bf 2021}.
\bibitem[Krawczyk et al. (2021)]{manga}
Krawczyk, C., Masters, K. Lingard, T. et al., MaNGA Morphologies from Galaxy Zoo, 
https://www.sdss.org/dr16/data\_access/value-added-catalogs/?vac\_id=manga-morphologies-from-galaxy-zoo
\bibitem[Owen \& Rudnick (1976)]{WAT}
Owen, F., Rudnick, L., Radio sources with wide-angle tails in Abell clusters of galaxies. {\em ApJ}, {\bf 1976}, {\em 205}, 1
\bibitem[Williams et al. (2019)]{lzoo}
Williams, W., Hardcastle, M., Best, P.N. et al., The LOFAR Two-metre Sky Survey. III. First data release: Optical/infrared identifications and value-added catalogue, {\em A\&A}, {\bf 2019}, {\em 622}, 2.

\end{thebibliography}

\end{paracol}
\end{document}